# Oscillating cosmological force modifies Newtonian dynamics

Igor I. Smolyaninov

*Department of Electrical and Computer Engineering, University of Maryland, College Park, MD 20742, USA*

**In the Newtonian limit of general relativity force acting on a test mass in a central gravitational field is conventionally defined by the attractive Newtonian gravity (inverse square) term plus a small repulsive cosmological force, which is proportional to the slow acceleration of the universe expansion. In this paper we consider the cosmological force correction due to fast quantum oscillations of the universe scale factor, which were suggested recently by Wang *et al*. (Phys. Rev. D 95, 103504 (2017)) as a potential solution of the cosmological constant problem. These fast fluctuations of the cosmological scale factor induce strong changes to the current sign and magnitude of the average cosmological force, thus making it one of the potential probable causes of the modification of Newtonian dynamics in galaxy-scale systems. The modified cosmological force may be responsible for the recently discovered "cosmic clock" behaviour of disk galaxies in the low redshift universe.**

The discrepancy between visible masses in galaxies and galaxy clusters and their dynamics predicted by the application of Newton's laws was first noticed by Fritz Zwicky almost hundred years ago [1]. The proposed solutions to this mystery range from dark matter halos around the galaxy centers [2] to various modifications of Newton's laws themselves [3,4] and even optical cloaking [5]. However, despite almost hundred years of experimental and theoretical studies, the ultimate solution of this mystery remains elusive [6]. The recently discovered "cosmic clock" behaviour [7] of disk galaxies in the low redshift universe further underlines this discrepancy. It appears that despite spanning a factor of 30 in size and velocity (from small irregular dwarf galaxies to the largest spirals) the galaxies behave as "clocks", rotating roughly once a billion years at the very outskirts of their discs. While this behaviour strongly contradicts conventional Newtonian dynamics in the absence of dark matter halos, it would also require a very special fine tuning of the dark matter halos around each galaxy.

On the other hand, it is well established for quite a while [8,9] that in an expanding universe the equations of motion of a test body in a central gravitational field must be modified even in the Newtonian limit of general relativity. The easiest way to see it is to write down the metric that describes the spacetime in the vicinity of a point mass $M$ placed in an expanding flat background [10]:

$$ds^2 = \left(1 - \frac{2GM}{a(t)\rho}\right)dt^2 - a(t)^2\left(d\rho^2 + \rho^2(d\theta^2 + \sin^2\theta d\phi^2)\right), \qquad (1)$$

where $G$ is the gravitational constant, $a(t)$ is the cosmological scale factor, and $\rho$ is the comoving radial coordinate, which relates to the proper radial coordinate $r$ as

$$r = a(t)\rho \qquad (2)$$

The geodesics corresponding to the line element (1) are defined by

$$-\left(\ddot{r} - \frac{\ddot{a}}{a}r\right) - \frac{GM}{r^2} + r\dot{\phi}^2 = 0 \tag{3}$$

Introducing $L$ as the constant angular momentum per unit mass

$$L = r^2\dot{\phi}, \tag{4}$$

the radial equation of motion for a test body in the Newtonian limit may be written as

$$\ddot{r} = \frac{\ddot{a}}{a}r - \frac{GM}{r^2} + \frac{L^2}{r^3} \tag{5}$$

As we can see from Eq.(5), the Newtonian equation of motion in an expanding universe is modified by an additional small cosmological force [8,9]

$$F_{\cos m} = m\frac{\ddot{a}}{a}r, \tag{6}$$

where $m$ is the mass of the test body. The current value of $\ddot{a}$ is believed to be positive based on the experimental measurements of the universe deceleration parameter

$$q = -\frac{\ddot{a}a}{\dot{a}^2} = -\frac{\dot{H}}{H^2} - 1, \tag{7}$$

where $H$ is the Hubble parameter. According to the Planck spacecraft data, in the current epoch $q \approx -0.55$ [11], which results in a small repulsive addition to the attractive Newtonian gravity term in eq.(5):

$$\frac{\ddot{a}}{a}r \approx 0.55H^2 r \tag{8}$$

At large distances from the center, the velocity distribution of the test bodies would be modified as

$$v = r\dot{\phi} = \sqrt{\frac{GM}{r} - 0.55H^2 r^2}, \tag{9}$$

leading to an apparent upper limit on the radius $R_M$ of a gravitationally bound system at a given M:

$$R_M = \left(\frac{GM}{0.55H^2}\right)^{1/3} \qquad (10)$$

Assuming the mass of the Milky Way to be M~$10^{12}$ solar masses, the projected $R_M$ for the Milky Way should be about $3 \times 10^6$ light years, or about 30 times larger than its radius. This should mean that the so determined cosmological force plays very little role in the Milky Way dynamics, and it cannot be responsible for the observed flattening of the rotation curves in a typical galaxy [2].

Fortunately, this may not be the end of the story for the cosmological force. In a very recent development Wang *et al.* suggested that very fast quantum fluctuations of the universe scale factor $a(t,\vec{x})$ may considerably alter the universe dynamics, thus providing a potential solution to the cosmological constant problem [12]. Let us demonstrate that these fast fluctuations of the cosmological scale factor also induce strong changes to the current sign and magnitude of the average cosmological force. As a result, the re-defined averaged cosmological force may contribute substantially to the modification of Newtonian dynamics in galaxy-scale systems.

Wang *et al.* [12] assumed the global metric of the universe to have the cosmology's standard Friedmann-Lemaître-Robertson-Walker (FLRW) form

$$ds^2 = dt^2 - a(t,\vec{x})^2\left(dx^2 + dy^2 + dz^2\right), \qquad (11)$$

while allowing spatio-temporal inhomogeneity in the scale factor $a(t,\vec{x})$. After solving the full coordinate-dependent Einstein field equations, they have obtained the following dynamic evolution equation for $a(t,\vec{x})$:

$$\ddot{a} + \omega^2(t, \vec{x}) a = 0, \tag{12}$$

where $\omega^2 > 0$ for quantum fluctuations of the matter fields (see Eqs.(41,42) from [12]). Due to the stochastic nature of these fast quantum fluctuations, $\omega(t, \vec{x})$ is not strictly periodic. However, as demonstrated in [12], its effect on the gravitating system is still similar to a periodic function (see also further refinements of these arguments in [13]). We should also note that cosmological solutions exhibiting fast scale factor oscillations appear quite generically in many other situations, such as semiclassical quantum gravity [14], cosmological models with oscillating dark energy [15], and various modified gravity theories [16]. In some of these situations (as for example in ref. [12]) the scale factor oscillations do not necessarily decay due to re-heating [17]. In addition, experimental evidence of relatively fast scale factor oscillations also started to emerge very recently [18]. Therefore, it makes sense to consider the effect of fast fluctuations of the cosmological scale factor on the sign and magnitude of the cosmological force described by Eqs.(5,6).

Following the standard analysis of system dynamics under the influence of a fast oscillating force [19], let us consider a toy cosmological model in which the evolution of the scale factor of the universe is separated into the slow and fast components:

$$a(t) = a_0(t) + \alpha \sin \omega t \tag{13}$$

In this decomposition $a_0(t)$ represents the observed slow cosmological evolution of the universe, while $\alpha(t) \ll a_0(t)$ represents the typical (potentially also time dependent) amplitude of the scale factor fluctuations introduced in [12] (in order to simplify our consideration we have neglected spatial dependence of $\alpha$ by replacing $\alpha(\vec{x})$ with its spatial average). We will assume that $\omega^{-1}$ is much faster than the typical time scales of galaxy evolution and the evolution of universe as a whole, since as suggested in [12], the physical origin of these fluctuations is due to quantum effects. We will also assume $\omega^{-1}$ to be much faster than any possible time dependence of $\alpha$, so its time derivative may be neglected. Under these assumptions the radial equation of motion of a test body in a central gravitational field will be modified as follows:

$$\ddot{r} = \frac{\ddot{a}_0}{a_0}r - \frac{\alpha\omega^2 \sin\omega t}{a_0}r - \frac{GM}{r^2} + \frac{L^2}{r^3}, \qquad (14)$$

which indicates that the cosmological force experiences fast oscillations. If these oscillations are fast enough, the $\alpha\omega^2$ term may not be neglected compared to the $\ddot{a}_0$ term. Following the standard treatment in [19], the proper radial coordinate of the test body should be expressed as

$$r = r_0(t) + \xi(t), \qquad (15)$$

where $\xi(t)$ represents small oscillations of the test body with respect to its slowly evolving radial position $r_0(t)$. By separating the fast and slow motion in Eq.(14), and by expanding to leading order in small quantities, the amplitude of these small oscillations may be defined from

$$\ddot{\xi} = -\frac{\alpha\omega^2 \sin\omega t}{a_0}r_0, \qquad (16)$$

thus leading to

$$\dot{\xi} = \frac{\alpha\omega \cos\omega t}{a_0}r_0 \qquad (17)$$

As usual (see [19]), we will assume that the average kinetic energy of the fast oscillations of the test body $\langle m\dot{\xi}^2/2\rangle$ contributes to its effective potential energy

$$U_{eff} = U + \frac{m}{2}\langle \dot{\xi}^2 \rangle \qquad (18)$$

7(similar to the introduction of the effective potential energy for an inverted (Kapitza) pendulum [19]), so that the average effective cosmological force in Eq.(5) must be re-defined as

$$\ddot{r}_0 = \frac{\ddot{a}_0}{a_0}r_0 - \frac{\alpha^2\omega^2}{2a_0^2}r_0 - \frac{GM}{r_0^2} + \frac{L^2}{r_0^3} \qquad (19)$$

Note that the newly derived component of the cosmological force (due to the fast scale factor fluctuations) is always attractive. The total overall cosmological force will also become attractive if

$$\frac{2\ddot{a}_0 a_0}{c^2} \approx \frac{1.05 H^2 a_0^2}{c^2} < \frac{\alpha^2\omega^2}{c^2} \qquad (20)$$

If the origin of the fast scale factor fluctuations is quantum mechanical, and both their characteristic amplitude $\alpha$ and frequency $\omega$ are defined by the Planck scale, the right hand side of Eq.(20) is expected to be of the order of unity. The left hand side of Eq.(20) has the same order of magnitude, which means that it is plausible that the overall sign of the cosmological force is attractive. Therefore, let us assume that

$$\ddot{r}_0 = -\frac{Ac^2}{a_0^2}r_0 - \frac{GM}{r_0^2} + \frac{L^2}{r_0^3} \quad , \qquad (21)$$

where $A\sim1$ is a dimensionless positive constant, which exact magnitude needs to be determined either from the future theory of quantum gravity, or from the astronomical observations. Such an attractive cosmological force will result in the following velocity distribution of the test bodies far from the compact central mass:

$$v = r\dot{\phi} = \sqrt{\frac{GM}{r_0} + \frac{Ac^2}{a_0^2}r_0^2} \quad , \qquad (22)$$

and at very large distances from the center a slow linear velocity increase will be observed as a function of radial distance:

$$v \propto A^{1/2} \frac{c}{a_0} r_0 \qquad (23)$$

As illustrated in Fig. 1, the measured rotation curve in such typical spiral galaxy as M33 indeed shows slow linear increase of rotation velocity with the radial distance. This behaviour is consistent with the expectations based on Eqs.(22,23) for $A^{1/2}$~0.05. The observed rotation curve of the Milky Way [20] (also shown in Fig.1) exhibits similar linear dependence of the rotation speed at large distances from the center.

The difference between the measured rotation curve and the rotation curve calculated based on the Newtonian dynamics of visible matter is typically attributed to a spherical dark matter halo [2,6], which is assumed to exist around virtually every galaxy. If such a spherical halo has approximately constant density, its mass distribution would be

$$M(r) = \frac{M_{halo} r^3}{R_{halo}^3}, \qquad (24)$$

leading to similar linear dependences of the gravitational and cosmological forces on the radial coordinate. This means that the effect of the dark matter halo and the effect of the cosmological force may mimic each other, and that the total amount of dark matter in a typical galaxy needs to be carefully re-examined. A good strategy to differentiate between the effects of dark matter halo and the cosmological force is to carefully re-evaluate the galaxy rotation curves at very large distances from the galactic centers. If these rotation curves show signs of a universal linear increase far from the galaxy center (see Eq.(23)), such an effect may turn out to become a very important observational evidence of quantum gravity. Comparison of the rotation curves at large distances from the center for the M33 and for the Milky Way galaxy shown in Fig,1 indicates that this research direction may indeed be very promising. The slopes of the rotation curves look nearly identical at large distances from the respective galactic centers.

In fact, recent detailed observations indicate that this is not a mere coincidence. It appears that disk galaxies in the low redshift universe exhibit the "cosmic clock" behaviour [7] illustrated in Fig.2. Despite spanning a factor of 30 in size and velocity (from small irregular dwarf galaxies to the largest spirals) the disk galaxies behave as "clocks", rotating roughly once a billion years at the very outskirts of their discs. While this behaviour strongly contradicts conventional Newtonian dynamics in the absence of dark matter halos, it would also require a very special fine tuning of the dark matter halos around each galaxy (see [21] as an example). It has to be assumed that the mass and the angular momentum of the visible matter disks must be fixed fractions of those of its surrounding dark matter halos. On the other hand, the cosmic clock behaviour finds very simple and natural explanation if the universal attractive cosmological force defined by Eq.(21) is responsible for the observed galaxy rotation curves at large distances from the centers. The universal linear long-distance rotation curve defined by Eq.(23) leads to universal "clock-like" periodicity of galactic rotations:

$$T = \frac{2\pi r_0}{v} = \frac{2\pi a_0}{cA^{1/2}}, \qquad (25)$$

which does not depend on the galaxy shape or dimensions. Indeed, low scatter around the "universal slope" $dv/dr=1.84 \times 10^{-3}$ km/s/ly seen in Fig.2 implies that this relationship is generic to disc galaxies in the low redshift universe. As a result, the galaxies behave as "cosmic clocks", rotating with roughly one billion years periodicity.

The magnitude of the "universal" $dv/dr$ stemming from the data in Fig.2 indicates that the true value of $A^{1/2}$ in Eq. (23) may be about twice as large as the $A^{1/2}$~0.05 value indicated by the data shown in Fig. 1. In order to clarify this issue and re-examine the assumed amount of dark matter in the M33 galaxy we have performed numerical simulations of the rotation curve in M33 assuming a disk-shape mass distribution of the visible matter, different magnitudes of $A^{1/2}$ ranging from 0 to 0.12, and without any contribution from the dark matter halo. Results of these simulations are shown in Fig.3. Numerical simulations performed using $A^{1/2}$=0.10 and $A^{1/2}$=0.12 show considerably improved agreement with the observational data (compare these results with Fig. 1: the huge difference between the observed and the calculated rotational velocities is almost eliminated). While such simplified model calculations cannot produce a perfect match with the observational data, this improved agreement is nevertheless quite remarkable. Note that these particular magnitudes of $A^{1/2}$ were suggested by the "cosmological clock" results shown in Fig.2. Our results strongly indicate that the typically assumed amounts of dark matter in disk galaxies must be carefully re-examined.

To summarize, based on the potential solution to the cosmological constant problem recently suggested in [12,13], which assumes fast quantum mechanical oscillations of the cosmological scale parameter, we have re-examined the sign and the magnitude of the cosmological force which modifies Newtonian dynamics around a central gravitating body in an expanding universe. The cosmological force in this model appears to be attractive, which means that it can mimic the effects of a spherical dark matter halo around a galaxy center. Therefore, the total amount of dark matter (if any) in a typical galaxy needs to be carefully re-examined by taking into account the effect of the cosmological force.

Since the long-distance behaviour of the cosmological force is expected to be universal, this mechanism provides simple and natural explanation of the "cosmic clock" behaviour [7] observed in the low redshift disk galaxies. If confirmed by further observations, this effect may turn out to become a very important observational evidence of macroscopic quantum gravity.

Finally, since sign of the cosmological force due to scale factor fluctuations appears to be always attractive, the popular Big Rip scenarios [8] will also need to be re-examined.

**References**

[1] F. Zwicky, "Die Rotverschiebung von extragalaktischen Nebeln", *Helvetica Physica Acta* **6**, 110–127 (1933).

[2] E. Corbelli, P. Salucci, "The extended rotation curve and the dark matter halo of M33", *Monthly Notices of the Royal Astronomical Society* **311**, 441-447 (2000).

[3] M. Milgrom, "A modification of the Newtonian dynamics as a possible alternative to the hidden mass hypothesis", *Astrophysical Journal* **270**, 365–370 (1983).

[4] J. D. Bekenstein, "Relativistic gravitation theory for the MOND paradigm", *Phys. Rev. D* **70**, 83509 (2004).

[5] I. I. Smolyaninov, "Galactic optical cloaking of visible baryonic matter", *Phys. Rev. D* **97**, 104008 (2018).

[6] S. Hossenfelder, S. McGaugh, "Is dark matter real?" *Scientific American*. **319**, 36–43 (2018).

[7] G. R. Meurer, D. Obreschkow, O. I. Wong, Z. Zheng, F. M. Audcent-Ross, D. J. Hanish, "Cosmic clocks: a tight radius-velocity relationship for HI-selected galaxies", *MNRAS* **476**, 1624-1636 (2018).

[8] S. Nesseris, L. Perivolaropoulos, "The fate of bound systems in phantom and quintessence cosmologies", *Phys. Rev. D* **70**, 123529 (2004).

[9] R. Nandra, A. N. Lasenby, M. P. Hobson, "The effect of a massive object on an expanding universe", *Monthly Notices of the Royal Astronomical Society* **422**, 2931–2944 (2012).

[10] A. Einstein, E. G. Straus, "The influence of the expansion of space on the gravitation fields surrounding the individual stars", *Rev. Mod. Phys*. **17**, 120 (1945).

[11] Planck Collaboration: P. A. R. Ade, *et al*. "Planck 2013 results. XVI. Cosmological parameters", *Astronomy & Astrophysics* **571**, A16 (2014).

[12] Q. Wang, Z. Zhu, W. G. Unruh, "How the huge energy of quantum vacuum gravitates to drive the slow accelerating expansion of the Universe", *Phys. Rev. D* **95**, 103504 (2017).

[13] S. S. Cree, T. M. Davis, T. C. Ralph, Q. Wang, Z. Zhu, W. G. Unruh, "Can the fluctuations of the quantum vacuum solve the cosmological constant problem?", *Phys. Rev. D* **98**, 063506 (2018).

[14] H. Matsui, N. Watamura, "Quantum spacetime instability and breakdown of semiclassical gravity", arXiv:1910.02186 [gr-qc]

[15] F. Pace, C. Fedeli, L. Moscardini, M. Bartelmann, "Structure formation in cosmologies with oscillating dark energy", *Mon. Not. R. Astron. Soc*. **422**, 1186–1202 (2012).

[16] F. Zaripov, "Oscillating cosmological solutions in the modified theory of induced gravity", *Advances in Astronomy* **2019**, 1502453 (2019).

[17] K. D. Lozanov, "Lectures on reheating after inflation", arXiv:1907.04402 [astro-ph.CO].

[18] H. I. Ringermacher, L. R. Mead, "Observation of discrete oscillations in a model-independent plot of cosmological scale factor versus lookback time and scalar field model", *The Astronomical Journal* **149**, 137 (2015).

[19] L. D. Landau and E. M. Lifshitz, *Mechanics*, 3rd ed. (Butterworth-Heinemann, Oxford, 1976), p. 131.

[20] J. O. Bennett, M. O. Donahue, N. Schneider, M. Voit. *The Essential Cosmic Perspective*, 8th ed. (Pearson, 2018).

[21] H. J. Mo, S. Mao, S. D.M. White, "The formation of galactic disks", *Mon. Not. Roy. Astron. Soc.* **295**, 319 (1998).

4**Figure Captions**

**Figure 1**. Comparison of the observed rotation curve (red) with the computed rotation curve (blue) obtained based on the unmodified Newtonian dynamics and the visible mass distribution for the typical spiral galaxy M33 [2]. Note that at large distances from the galactic center the observed rotation curve exhibits slow linear dependence on radial distance, which is consistent with the expectations based on Eqs.(22,23) for $A^{1/2} \sim 0.05$. The observed rotation curve of the Milky Way (black) [20] exhibits similar linear dependence of the rotation speed at large distances from the center.

**Figure 2**. Log-log plot of the observed circular velocity $V$ as a function of maximum radius $R$ for low redshift disk galaxies re-plotted using data from Fig.4 of ref. [7]. Low scatter around the "universal slope" $1.84 \times 10^{-3}$ km/s /ly indicated by the red line implies that this relationship is generic to disc galaxies in the low redshift universe. As a result, the galaxies behave as "cosmic clocks", rotating roughly once a billion years at the very outskirts of their discs. This "universal slope" corresponds to $A^{1/2} \sim 0.10$.

**Figure 3**. Numerical simulations of the rotation curve in M33, which were performed for different magnitudes of $A^{1/2}$ between 0 and 0.12 assuming a disk-shape mass distribution of the visible matter and no dark matter halo. Calculations performed using $A^{1/2}=0.10$ and $A^{1/2}=0.12$ show considerably improved agreement with the astronomical data (compare these results with Fig. 1: the huge difference between the observed and the calculated rotational velocities is almost eliminated).

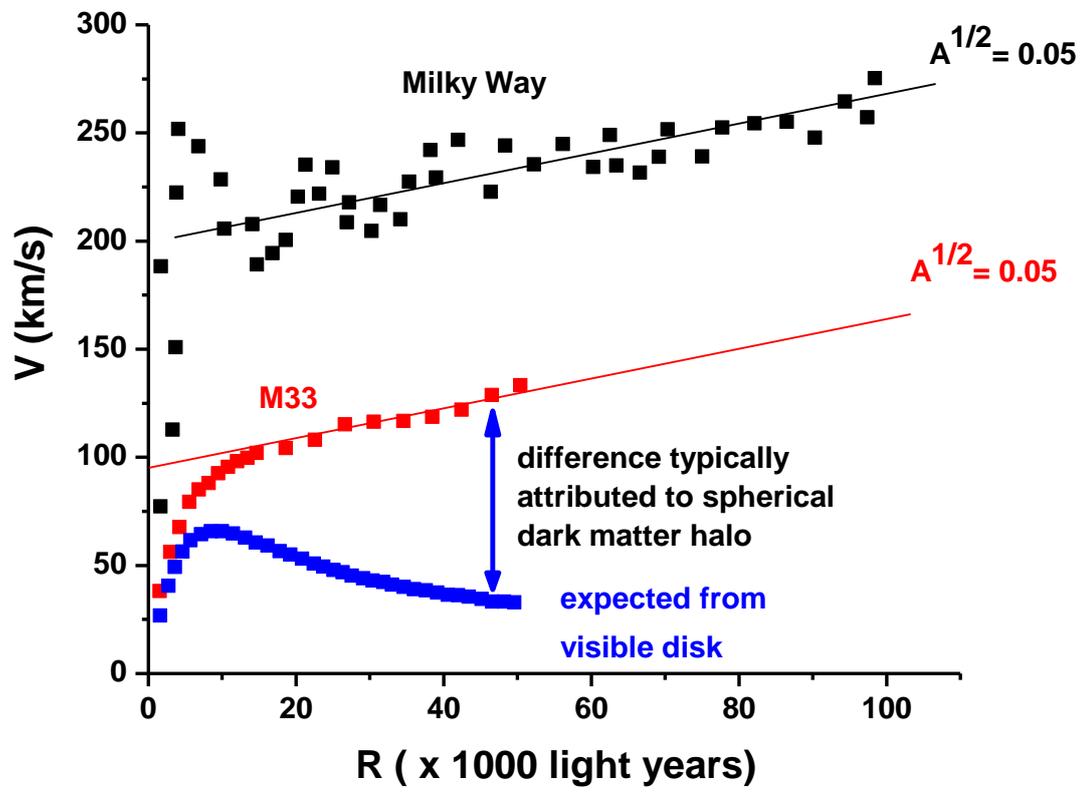

Fig. 1

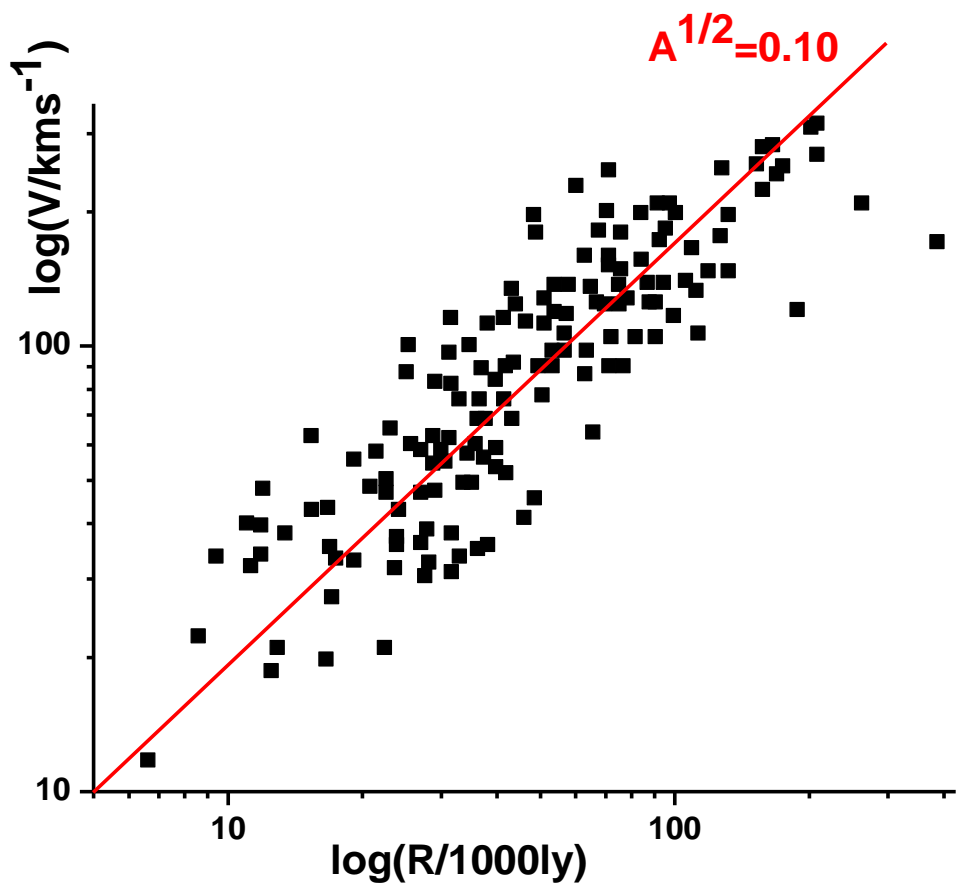

Fig. 2

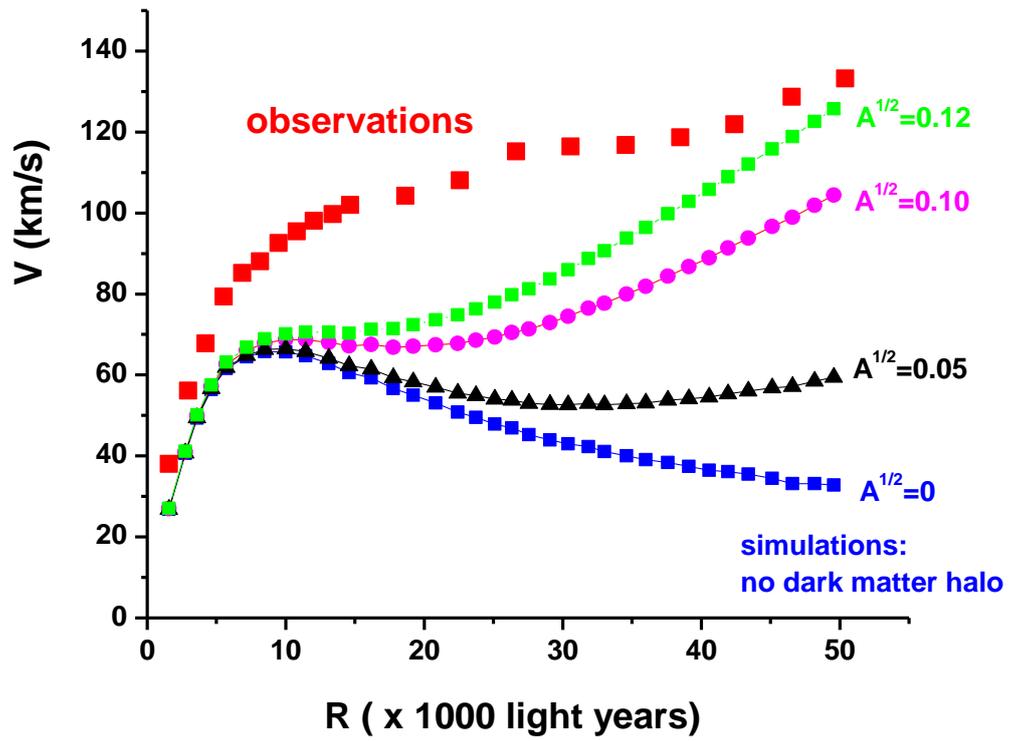

Fig. 3